\documentstyle[12pt,epsfig]{article}
\newcommand{\be}{\begin{equation}}
\newcommand{\ee}{\end{equation}}
\newcommand{\bea}{\begin{eqnarray}}
\newcommand{\eea}{\end{eqnarray}}
\newcommand{\beas}{\begin{eqnarray*}}
\newcommand{\eeas}{\end{eqnarray*}}
\newcommand{\bi}{\begin{itemize}}
\newcommand{\ei}{\end{itemize}}
\newcommand{\bc}{\begin{center}}
\newcommand{\ec}{\end{center}}
\newcommand{\bfl}{\begin{flushleft}}
\newcommand{\efl}{\end{flushleft}}
\newcommand{\bfr}{\begin{flushright}}
\newcommand{\efr}{\end{flushright}}
\newcommand{\f}{\frac}

\def\6{\partial} \def\a{\alpha}

\def\r{\rho} \def\s{\sigma}

\newcommand{\ZZ}{{\cal Z}}
\begin{document}
\title{Note on Four $Dp$-Branes at Angles}
\author{Ion V. Vancea
\footnote{On leave from Babes-Bolyai University of Cluj, Romania}\\
{\em Instituto de F\'{\i}sica T\'{e}orica, Universidade Estadual Paulista} \\
{\em Rua Pamplona 145, 01405-900 S\~{a}o Paulo SP, Brazil}\\
and\\
{\em Grupo de F\'{\i}sica T\'{e}orica, Universidade Catolica de Petr\'{o}polis}\\
{\em Rua Bar\~{a}o do Amazonas 124, 25685-000 Petr\'{o}polis RJ, Brazil}}
\maketitle

\begin{abstract}
In this note we analyse the dynamical potential of a system of four 
$Dp$-branes at arbitrary angles. The equilibrium configurations for various 
values of the relative angles and distances among branes are discussed. The
known configurations of parallel branes and brane-antibranes are obtained
at extrema of the dynamical potential.

\end{abstract}

\setcounter{page}{0}
\thispagestyle{empty}

\newpage

\section{Introduction}
The $D$-branes at angles, viewed either as boundary states in the Fock space 
of the closed strings \cite{asj,sj,aa} or as solitons in the low energy limit 
of string theories \cite{ot,ts,pkt,bdl} can form interesting systems. 
Recently, they had been realized in super Yang-Mills theories 
\cite{koo,oz,koz,aat} and it was shown in \cite{afiru,aiqu,au,fhs,bgk} that 
the cancellation of the tadpole anomaly in type II theories compactified on 
$Z_N \times Z_M$ orbifolds requires the introduction of $D$-branes at angles
in order to produce supersymmetric non-chiral field theories. On the other
hand, there were found static solutions to Einstein's equations corresponding 
to branes at angles that intersect on a three-brane in the context of the 
Randall-Sundrum model \cite{cs}. A deficit angle in the transverse space
of branes was used in a tentative to motivate a critical cosmological constant
\cite{tt}. Also, the branes at angles were employed in the modelling of 
black-holes \cite{vb}.  Branes at angles on compact manifolds and in 
the Born-Infeld field theory were studied in \cite{ggpt,afs,dk,clp}. More
recently, their connection with the noncommutative geometry has been 
investigated in \cite{wy,cimm,hh}. 

The interacting potential between two $Dp$-branes depends on their relative 
angles \cite{bs,am,ms} and it has relative and absolute extrema which describe 
a brane-antibrane system or a configuration of two parallel branes, 
respectively. Using several $D$-branes and $NS$-branes, some stable 
configurations of branes-antibranes can be obtained \cite{smnvs,smnvsdt} by
compensating geometrically the interaction generated by the tachyonic fields
\cite{spa}. However, it is interesting to see if there are any stable 
configurations of $D$-branes only. Intuitively, one would say that it is 
possible to find some values of the relative angles between branes for which 
the potential of the system reaches an extrema. The aim of this short note is
to address the question of stability in the case of a system of four 
D$p$-branes at angles. This system is more richer than a system of three 
branes since it includes the latter one and presents a configuration of 
two brane-antibrane pairs.

The outline of the paper is as follows. In Section 2 we review the basic
features of a system formed by two branes at angles. In Section 3 we
construct the dynamical potential of four branes and analyse their 
configurations.
In Section 4 we present some configurations that describe brane-antibranes.
In these two sections the effects due to the presence of the tachyons are
ignored. The reason is that the tachyons should be described by an
open string field theory and at present we do not know any such of theory that
describes branes at angles (but see \cite{ivv}). However, the tachyons play 
an important role in the dynamics and stability. Therefore, in Section 5 some 
general comments on the tachyons of the system are made. 
The last section is devoted to discussions. 

\section{Two branes at angles}

Let us consider two $Dp$-branes in type II string theories that make a 
relative angle $\theta$ in the
$(p,p+1)$ plane and are separated by a vector $Z^{\mu}$. Consider an open 
string stretched between the branes with one end at $\sigma =0$ on one
brane and the other end at $\sigma = \pi$ on the other brane. The boundary 
condition are the usual ones on the directions outside this plane and 
are given by 
\bea
X^{(p)}\sin\theta-X^{(p+1)}\cos\theta=0\nonumber\\
\6_{\sigma}X^{(p)}\cos\theta+\6_{\sigma}X^{(p+1)}\sin\theta=0
\label{boundcondpi}
\eea
in the plane, on the brane on which $\sigma =\pi$. The fermionic boundary 
conditions follow from the requirement that the string be supersymmetric.
In the RNS formalism in superconformal gauge they are given by the usual
boundary conditions outside the $(p,p+1)$-plane and by
\bea
\bar{\epsilon}\r^1\r^0\psi^{(p)} &=& 0 \nonumber\\
\bar{\epsilon}\r^1\r^1\psi^{(p+1)} &=& 0, 
\label{boundcondferm0}
\eea
for $\s = 0$ and
\bea
\bar{\epsilon}\r^1\r^0(\cos\theta\psi^{(p)} +
\sin\theta\psi^{(p+1)}) &=& 0 \nonumber\\
\bar{\epsilon}\r^1\r^1(\cos\theta\psi^{(p)} -
\sin\theta\psi^{(p+1)}) &=& 0,
\label{boundcondfermpi}
\eea
for $\s=\pi$.
Here, $\r^0$ and $\r^1$ are the complex Dirac matrices and $\epsilon$ is 
an arbitrary two-dimensional Majorana spinor that parametrizes the 
supersymmetry transformation.
The system can be quantized in the canonical formalism \cite{asj,sj,aa}
and the usual Ramond and Neveu-Schwarz sectors are obtained with the ranges 
of the indices of the Fourier modes in $\ZZ + \theta/\pi$ in $R$ sector and 
$\ZZ + 1/2 + \theta/\pi$ in the $NS$ sector.

The interaction between the branes can be calculated from the exchange
of states from the closed string channel. The scattering amplitude of the 
massless states ($\rightarrow 0$ limit) is given in terms of the dual open 
string variables by the following relation
\cite{asj}
\be
A(\theta,Z) = V_p \int \f{dt}{t}(8\pi^2 \a 't)^{-\f{p}{2}}
e^{-\f{Z^2t}{2\pi^2\a '}}[8t^3\tan(\f{\theta}{2})\sin^2(\f{\theta}{2})] 
\label{twobramplit}
\ee
and it is computed as in the $\theta = 0$ case \cite{jp}.

The dynamical potential of the long range interactions has contributions from 
both 
$R$ and $NS$ sectors and its form can be read off Eq.(\ref{twobramplit}).
If we
consider for simplicity that $Z^{\mu}$ has just one component different from
zero then the potential has the following form
\be
V(\theta,Z) = -V_p4(4\pi\a ')^{3-p}\pi^{\frac{p}{2}-4}\Gamma(3-\f{p}{2})
Z^{p-6}\f{(1-\cos\theta)^2}{\sin\theta}.
\label{twobrpot}
\ee
For $\theta \in [0,\pi]$ the potential above has an absolute maximum at
$\pi$ where it blows up. This is interpreted as an unstable brane-antibrane
configuration which eventually collapses to an brane of lower dimension. The
reson for that is the presence of a tachyon that is not removed from the
spectrum by the GSO projection \cite{as}. Let us remark that one can extend
the analysis for the case when $\theta \in [0,2\pi]$. The boundary conditions
(\ref{boundcondpi}) remain the same if we replace the angle $\theta$ by
$\theta + \pi$. Therefore, the solution of the bosonic equations of motion is
the same and by supersymmetry we will obtain the same spinorial solutions.
However, if we plot the potential for the full interval $[0,2\pi]$, we see
that $\pi$ is in the same time an absolute minimum of the function
(see Fig.(\ref{figpotang})).

\newcommand{\postscript}[2]
 {\setlength{\epsfxsize} {#2\hsize} \centerline {\epsfbox{#1}}}

\begin{figure}
\postscript{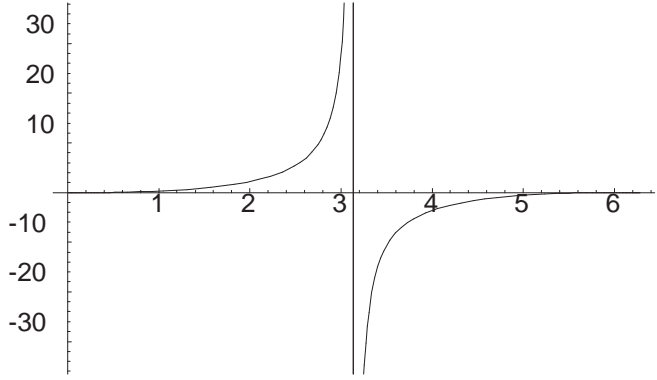}{0.5}
\caption{The interacting potential between two $D$p-branes in the
interval $[0,2\pi]$ with a fixed value of
the constants and for an arbitrary finite separation between branes.}
\label{figpotang}
\end{figure}

One possible interpretation is that for small angles above $\pi$ the system
will tend to assume the brane-antibrane configuration and at this point the
system decays to a stable brane of lower dimension \cite{as}. At $0$ and 
$2\pi$ the systems has a zero potential. This corresponds to two parallel 
branes between which the $NS$ and $R$ contributions to potential cancell each
other. 

The system breaks all the supersymmetries of the background for arbitrary
angle between branes \cite{jpb}. In the non-supersymmetric configurations
a tachyon appears for any value of the angle between branes at distances 
below some critical value. 

\section{Dynamical potential of four branes at angles}

In this section we will discuss the dynamical potential of a system of four
$Dp$-branes. This potential is obtained by integrating over the amplitude of
exchange massless modes of closed strings.  

Let us consider a system of four $Dp$-branes that make one relative angle
between any of two of them in the $(p, p+1)$ plane. We denote the four branes
be $p$, $\bar{p}$, $p'$ and $\bar{p'}$, respectively. In what follows we will
consider only two brane processes in closed string tree level approximation.
The sectors entering in this interaction and the general configuration of the 
system are parametrized by three angles and three relative distances which are
chosen accordingly to the Table(\ref{Param}). 
\bc
\begin{tabular}{||c|c|c||}                \hline\hline
$Pair$             &  $Angles$               & $Distance$    \\\hline\hline
$p-p'$             &  $\phi$                 & $Y$           \\
$p-\bar{p}$        &  $\omega $              & $L_1$         \\
$p-\bar{p'}$       &  $\phi + \chi$          & $Y+L_2$       \\
$p'-\bar{p}$       &  $\phi - \omega$        & $Y-L_1$       \\
$p'-\bar{p'}$      &  $\chi$                 & $L_2$         \\ 
$\bar{p}-\bar{p'}$ &  $\phi + \chi - \omega$ & $Y+L_2 - L_1$ \\\hline
\end{tabular}
\label{Param}
\ec
The strings stretching between any two branes have boundary conditions of the
type (\ref{boundcondpi}) in the $(p, p+1)$ plane. The interactions between 
branes are superpositions of two brane interactions. Therefore, the total 
potential is just a sum of the potentials from all
sectors above 
\be
V(\theta_i,L_i) = \sum_i V_i(\theta_i,L_i),
\label{sumpot}
\ee
where $\theta_i \in \{\phi, \omega, \phi + \chi , \phi - \omega , \chi, 
\phi + \chi - \omega \}$ and 
$L_i \in \{ Y, L_1, Y+L_2, Y-L_1,L_2, Y+L_2-L_1 \}$. Explicitely, the 
dynamical potential
for long range interactions has the following form 
\bea
V & \sim & \f{(1-\cos\phi)^2}{\sin\phi}Y^{p-6} +
           \f{(1-\cos\omega)^2}{\sin\omega}L_1^{p-6} +
           \f{(1-\cos\chi)^2}{\sin\chi}L_2^{p-6}
           \nonumber\\
     & + & \f{(1-\cos(\phi+\chi))^2}{\sin(\phi+\chi)}(Y+L_2)^{p-6} +
           \f{(1-\cos(\phi-\omega))^2}{\sin(\phi-\omega)}(Y-L_1)^{p-6}
           \nonumber\\
     & + & \f{(1-\cos(\phi + \chi - \omega))^2}{\sin(\phi +\chi - \omega)}
           (Y+L_2 - L_1)^{p-6},
\label{explsumpot}
\eea
where $\sim$ means that we are considering the equality up to a numerical
constant.

We would like to know whether there are any values of the relative angles 
among branes that extremise the potential (\ref{explsumpot}). This is
equivalent to finding the solutions of the following system
\bea
F (\phi ) Y^{p-6} & + & F (\phi + \chi ) (Y+ L_2)^{p-6} + 
F (\phi - \omega )(Y- L_1 )^{p-6}
\nonumber\\
 &+& F (\phi + \chi - \omega ) (Y + L_2 - L_1 )^{p-6} = 0
\nonumber\\ 
F (\omega ) L_1^{p-6} & - & F (\phi - \omega )( Y- L_1 )^{p-6} - 
F (\phi + \chi - \omega)(Y + L_2 - L_1 )^{p-6} = 0
\nonumber\\
F (\chi )L_2^{p-6}  & + & F (\phi + \chi )(Y+L_2)^{p-6} +
 F (\phi + \chi - \omega )(Y + L_2 - L_1 )^{p-6} = 0,
\label{eqminang}
\eea
where  $F(\theta_i ) = (1-\cos\theta_i)(2+\cos\theta_i)/(1+\cos\theta_i)$. In
(\ref{eqminang}) the distances are considered fixed and finite. 

\subsection{Configurations in which the potential has an extrema}

In general, the solutions of the system (\ref{eqminang}) will depend on the 
parametes $Y, L_1, L_2$. However, as it is easy to see, there are some 
solutions independent of all distances. These describe configurations with 
arbitrary combinations of the following values of angles
\be
\phi = 0, 2\pi~~, ~~\omega = 0, 2\pi~~,~~\chi = 0,2\pi .
\label{solindpar} 
\ee
The result is known and it says that parallel branes form a stable system.
In these configurations the value of the potential is zero.

\subsection{$\phi = 0$}

More general solutions of (\ref{eqminang}) will depend on certain values of
the $Y$, $L_1$ and $L_2$. Due to the fact that the system above is
degenerate, we investigate some particular cases.

Let us assume that $\phi = 0$ which implies that $F(\phi )$ is also zero.
A nontrivial solution, i.e. a solution in which not all of the remaining
$G$'s are zero, can exist only if the following equation is satisfied 
\bea
(Y&+&L_2)^{p-6}[L_1^{p-6} - (Y-L_1)^{p-6}](Y+L_2-L_1)^{p-6}+
\nonumber\\
(Y&-&L_1)^{p-6}[L_2^{p-6} + (Y+L_2)^{p-6}](Y+L_2-L_1)^{p-6}
\nonumber\\
(Y&+&L_2 -L_1)^{p-6}[L_1^{p-6}-(Y-L_1)^{p-6}][L_2^{p-6}+(Y+L_2)^{p-6}]=0. 
\label{conddistone}
\eea
Once (\ref{conddistone}) solved, one can find, in priciple, the values of 
$F$'s 
for which the potential has an extrema, and from them one can deduce the
angles of the configuration, if any.  

An obvious solution of the equation (\ref{conddistone}) is given by $Y=L_1-L_2$
(we consider only positively definite distances.) 
A general solution is a root of the equation of degree $3(p-6)$ in one 
distance, say $Y$, in function of the other two. 
In the limit when $L_1 \rightarrow 0$, $L_2 \rightarrow 0$, the extrema can
be obtained only if $Y \rightarrow 0$, that is if the system collapses. In
the case of six branes the system is stable only if the angles $\chi$ and
$\omega$ are simultaneously zero or $2\pi$.
 
\subsection{$\phi = 0, \chi = \omega$}

To obtain nontrivial solutions in this case one has to set all the distances
to zero. The other possibility is to have the following relationship between
$F$'s satisfied
\be
F(\chi) = F(\omega)(\f{L_1}{L_2})^{p-6},
\label{soleqang}
\ee 
which holds only for $p \neq 6$. For $p=6$ the solution is given by 
(\ref{solindpar}). In addition, the distances $L_1$ and $L_2$ should satisfy
the following relation
\be
(2L_1 + L_2)^{p-6} = - L_2^{p-6},
\label{firstll}
\ee
while $Y = 2L_1$. For four, five and eight branes the configurations are 
easly red off these conditions.

\subsection{$\omega = 0$}

The system (\ref{eqminang}) is degenerate in this case, too. However, the
corresponding relation 
\bea
L_2^{p-6}[(Y & + & L2)^{p-6} + (Y+L_2 - L_1)^{p-6}] -
\nonumber\\
L_2^{p-6}( Y & + & L_2 - L_1)^{p-6}[Y^{p-6} + (Y-L_1)^{p-6}]=0
\label{conddisttwo}
\eea
admits solution for $p=6$ if $Y - L_1 \neq 0$ and $Y + L_2 - L_1 \neq 0$, in
contrast with the previous case (\ref{conddistone}). For any of the limits
$L_1 \rightarrow 0$ or $L_2 \rightarrow 0$, the relation cancells for any 
values of the remaining two parameters.

\subsection{$\chi = 0$}

The relation that resolve the degeneracy of the system is given by
\bea
L_1^{p-6}(Y & + & L_2)^{p-6}[(Y-L_1)^{p-6} + (Y+L_2 - L_1)^{p-6}]-
\nonumber\\
L_1^{p-6}(Y & + & L_2 - L_1)^{p-6}[Y^{p-6} + (Y+L_2)^{p-6}]=0.
\label{conddistthree}
\eea 
In this case, there are solutions for $p=6$ for any value of all parameters.
In the limit where $L_1 \rightarrow 0$, the potential can have an extrema 
for any $Y$ and $L_1$. At $L_2 \rightarrow 0$ we have solutions for arbitrary
$Y$ and $L_1$. If we set now the angles $\phi = \omega$ we see that the 
potential can have an extrema only if the system collapses.

We may ask what happens when the potential is varied with respect to the 
parameters
$Y$, $L_1$ and $L_2$, respectively. Since the derivative of the potential
with respect
to the spatial variable is the definition of the force, we may rephrase this
question
by writing the condition of stability of the system in terms of forces acting
on branes. In this situation, the system would be stable if each of the brane
is in equilibrium, that is if all forces acting of each brane cancell each 
other.

This condition can be casted into the following form for $p\neq 6$
\bea
&G(\omega)&L_1^{p-5} + G(\phi)Y^{p-5} + G(\phi + \chi)(Y+L_2)^{p-5}=0
\nonumber\\
-&G(\omega)&L_1^{p-5} + G(\phi -\omega)(Y-L_1)^{p-5} + G(\phi + \chi -\omega)
(Y+L_2 - L_1)^{p-5}=0
\nonumber\\
-&G(\phi)&Y^{p-5} - G(\phi - \omega)(Y-L_1)^{p-5} + G(\chi)L_2^{p-5}=0
\nonumber\\
-&G(\phi&+\chi)(Y+L_2)^{p-5} - G(\phi + \chi - \omega)(Y+L_2-L_1)^{p-5} -
G(\chi)L_2^{p-5}=0,
\label{forces}
\eea
where $G(\theta_i)$ is the term from the potential that depends on $\theta_i$.
The system above is degenerate, therefore the best one can do is to express 
three of the terms containing distances in terms of the other three, 
containing the angles. Again, particular solutions can be obtained by setting 
some parameters to constants. For example, for the angle $\phi=0$ and the 
other angles undetermined we obtain a relation between the spatial parameter
\be
\f{(Y+L_2)^{p-5}}{L_1^{p-5}} = 
\f{L_2^{p-5}}{(Y-L_1)^{p-5}} = 
\f{L_2^{p-5}+(Y+L_2)^{p-5}}{L_1^{p-5}+(Y-L_1)^{p-5}}=0.
\label{fordisone}
\ee
It is easy to see that for $L_2^2 + L_1^2 + 6L_1L_2 > 0$ there are acceptable
solution of (\ref{fordisone}) which give the expression of $Y$ in terms of 
$L_1$ and $L_2$.

In a similar way we can discuss the configurations at fixed angles 
$\omega = 0$ and $\chi = 0$. The correspondig relations are given by
\be
\f{(Y+L_2)^{p-5}}{Y^{p-5}}=
\f{(Y+L_2-L_1)^{p-5}}{(Y-L_1)^{p-5}}=
-\f{(Y+L_2-L_1)^{p-5} + (Y+L_2)^{p-5}}{Y^{p-5} + (Y-L_1)^{p-5}}
\label{fordistwo}
\ee
and
\be
\f{(Y-L_1)^{p-5}}{Y^{p-5}}=
\f{(Y+L_2-L_1)^{p-5}}{(Y+L_2)^{p-5}}=
\f{(Y+L_2-L_1)^{p-5} + (Y-L_1)^{p-5}}{Y^{p-5}+(Y+L_2)^{p-5}},
\label{fordisthree}
\ee
respectively. We consider in all relations above that the denominators do not
vanish.
The (\ref{fordisone}), (\ref{fordistwo}) and (\ref{fordisthree}) represent 
necessary conditions for the stability of the system. However, they should
not be compatibile to each other since they were established for different 
values of angles.

\section{Brane-antibrane pairs}    

For each relative angle that equals $\pi$ there is an antibrane in the system.
The non-equivalent configurations of branes-antibranes are the ones containing
one or two antibranes. For three antibranes the system is equivalent with a 
system with one antibrane.

\begin{figure}
\postscript{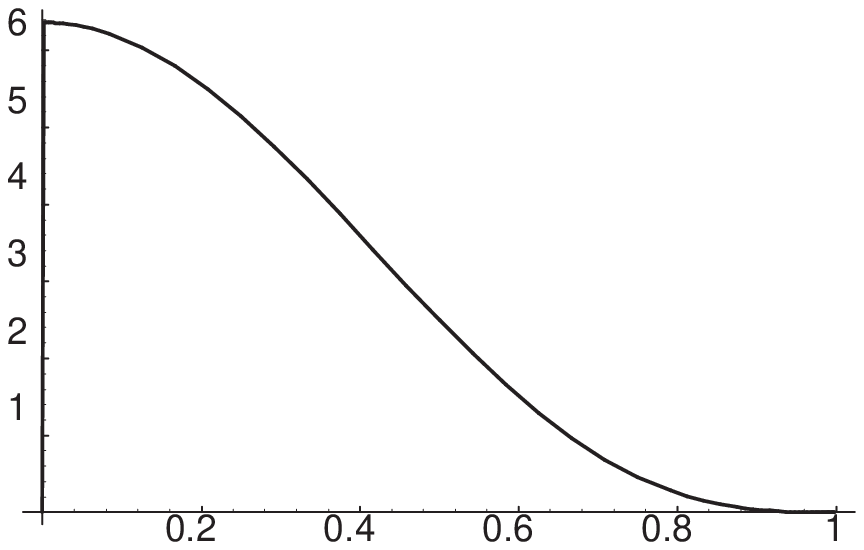}{0.5}
\caption{The behaviour of the dynamical potential of $D$7-branes
with one brane-antibrane pair in the interval $t\in [0,1]$ where 
$\omega  = (1-t)\pi$ and $L_1=tL_0$.}
\label{figpotf7}
\end{figure}

\begin{figure}
\postscript{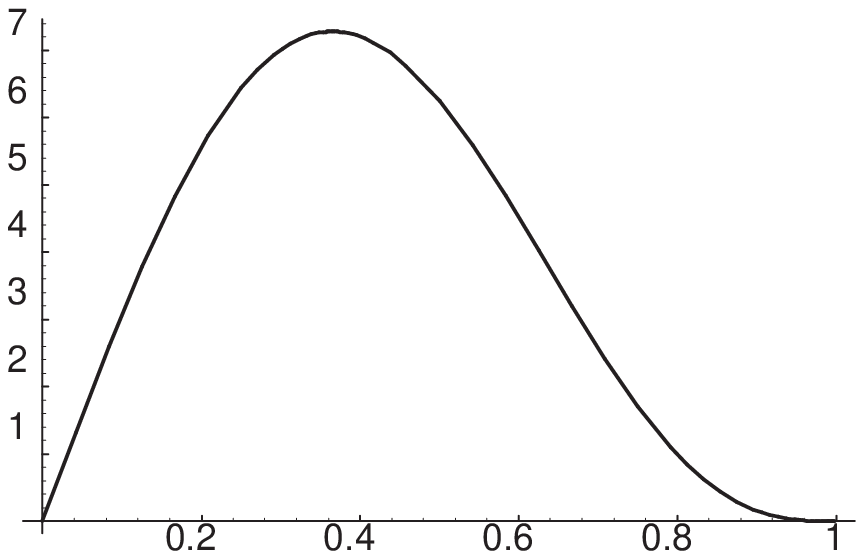}{0.5}
\caption{The behaviour of the dynamical potential of $D$8-branes
with one brane-antibrane pair in the interval $t\in [0,1]$ where 
$\omega = (1-t)\pi$ and $L_1=tL_0$.}
\label{figpotf8}
\end{figure}

\begin{figure}
\postscript{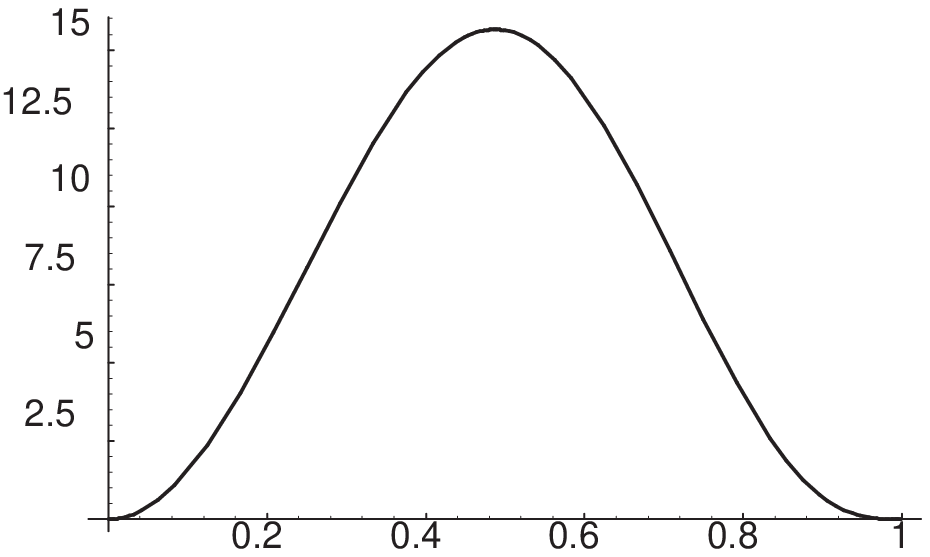}{0.5}
\caption{The behaviour of the dynamical potential of $D$9-branes
with one brane-antibrane pair in the interval $t\in [0,1]$ where 
$\omega = (1-t)\pi$ and $L_1=tL_0$.}
\label{figpotf9}
\end{figure}

\begin{figure}
\postscript{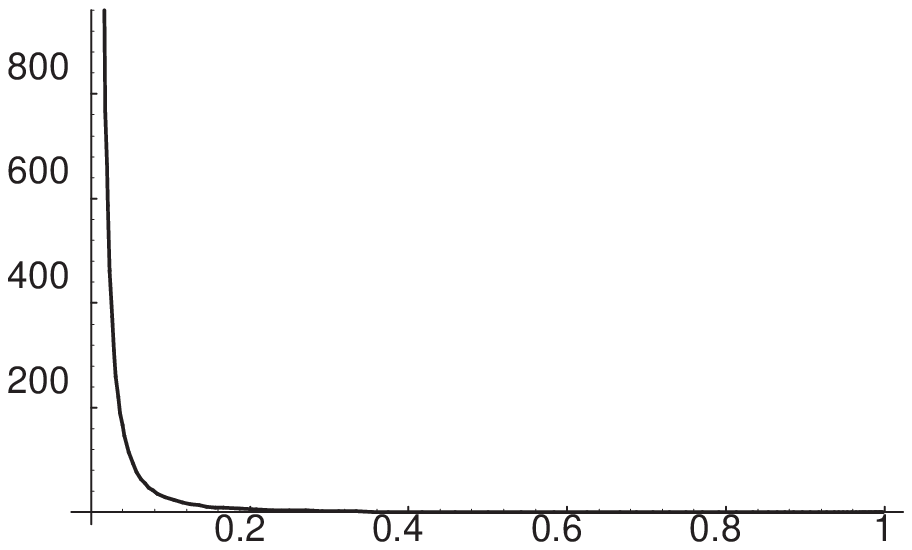}{0.5}
\caption{The typical behaviour of the dynamical potential of$D$p-branes,
$p\leq6$,
with one brane-antibrane pair in the interval $t\in [0,1]$ where 
$\omega = (1-t)\pi$ and $L_1=tL_0$.}
\label{figpotfp}
\end{figure}

One antibrane can be obtained by setting $\omega=0$. If $\phi$ and $\chi$ are
left arbitrary, the system will contain one brane-antibrane pair and two
branes at arbitrary angle. We assume that these angles do not equal $\pi$.
The potential $V(\phi, \omega = \pi, \chi )$ for any finite distances between
branes is as in Fig.(\ref{figpotang}). However, for certain behaviour of the 
parameters $\omega$ and $L_1$ the potential is finite as 
$\omega \rightarrow \pi$ and $L_1 \rightarrow 0$. To see this, we assume that
$L_1$ and $\omega$ vary towards $0$ and $\pi$, respectively, with the same
parameter $t$ which we pick up to be between $[0,1]$. The dependence of the
two parameters on $t$ should be $L_1 = tL_0$ and $\omega = (1-t)\pi$ 
where $L_0$ is
a constant. Then for $p=7, 8, 9$ the potential has a finite value at $t=0$
as shown in Fig.(\ref{figpotf7}),(\ref{figpotf8}) and (\ref{figpotf9}).
The typical behaviour of $p \leq 6$ branes is illustrated in the
Fig. (\ref{figpotfp}).
In plotting the potentials above, the distance $L_0$ was chosen positive and
greater than one. A special case is obtained when $\phi = \chi = 0$. The
configuration described in this case is that of three parallel branes and one 
antibrane among them. Thus there are three brane-antibrane pairs. The 
potential blows up in the neighbourhood of $\pi$ for any values of the
parameters $Y$, $L_1$ and $L_2$ least the following relation is satisfied
\be
(Y+L_2 - L_1)^{p-6} + (Y-L_1)^{p-6} - L_1^{p-6}=0
\label{condbab}
\ee
for which its value is  undetermined. We notice that when the brane-antibrane
are one a top of the other, the potential actually continues to be infinite.
If the two branes are on the top of the other, there are two real solutions
for $Y$ in terms of $L_1$: $Y_{1,2}=(2\pm \sqrt{3})L_1$.  

A configuration of two brane-antibrane pairs is obtained when two relative 
angles are set to $\pi$. If we choose $\omega=\chi = \pi$ and leave $\phi$
arbitrary, we see that, as in the previous case, the potential will diverge
in this configuration due to terms of the form $(1+1)^2/0$. The only 
possibility of making these terms finite is when their coefficients go 
to zero as the angles go to $\pi$. This implies that the following equation
should be satisfied by the relative distances between branes
\be
L_1^{p-6} + L_2^{p-6} + (Y+L_2)^{p-6} + (Y - L_1)^{p-6} = 0.
\label{condbabfin}
\ee
Again, branes of different dimensionality will allow different solutions
for Eq.(\ref{condbabfin}). For $p=6$ there is no possibility of getting a 
finite potential, while for $p=7$ one should set $Y=-L_2$ for any value of
$L_1$. 
 
\section{Effects of tachyons}

In the previous sections we have analysed the configurations of four 
$Dp$-branes that make relative {\em arbitrary} angles and are spaced by 
relative {\em arbitrary} distances and we discussed some general 
configurations given by the extrema of the dynamical potential. However,
it is known that for arbitrary values of angles and distances, a pair
of $Dp$-branes is in a non-supersymmetric configuration \cite{jpb}. In
this situation, since the $NS$ ground state of the system depends on the 
relative distance and angle between the branes its mass is given by
\cite{jpb}
\be
\a ' m^2 = \f{Z^2}{4\pi^2 \a '} + \f{\theta}{2\pi}.
\label{groundmass}
\ee
Thus, for any given angle $\theta \in [0,2\pi ]$ there
is a critical distance under which the ground state is a tachyon. The 
tachyon potential will affect the stability of the system. As was conjectured
in \cite{as} the system will evoluate until the potential will reach to 
a minimum.
For a brane anti-brane system, one on the top of the other, the minimum will 
be reached in a configuration that represents a $Dp$-brane of lower 
dimensionality. 

In the case of four $Dp$-branes at angles, the same reasoning applies to all
pairs and six tachyons may appear if the relative distances among branes are
smaller than the critical values. For $0 \leq \theta_i \leq \pi$ these are
given by the following relations:
\be
L^{2}_{ic} = (2\pi \a ')_1\theta_i
\label{criticaldist}.
\ee
The equations above impose lower bound limits for the validity of the results
obtained in the previous sections. If any of $L^{2}_i < L^{2}_{ic}$ then a
tachyon appears in the corresponding sector and the dynamics will be 
determined by the tachyon potential. However, the tachyon is an off-shell 
degree of freedom of string, and therefore its dynamics cannot be described
by a first quantized string theory. If some of the relative angles $\theta_a 
 =\pi$ and if the interaction among the tachyons is neglected,
the tachyon potential is given by \cite{bsz}
\be
V(T_a) = \sum_a e^{-\f{1}{4}T^{2}_a} = 1 - \sum_a \f{1}{4}T^{2}_{a} + \cdots,
\label{tachpot}
\ee
and the tachyons condensate at $T_a \rightarrow \infty$ leaving behind lower 
dimensional branes in that sectors. However, the value of the tachyonic 
potential is not know for general angles. We hope to be able to say more on 
this topic in a future work \cite{ivv}.

\section{Discussions}
\setcounter{equation}{0}

From the configurations analysed above, we can see that the extremum 
of the interacing potential between
four branes at angles can be expressed as a condition between relative
angles and distances among branes. For fixed angles, the condition reduced
to some relations between distances, while for fixed distances, the
extremum of the potential implies some relations between functions of cosine
of angles. In general the angular factors $F(\theta_i)$ and $G(\theta_i)$
cannot be determined uniquely from the degenerate systems in which they 
enter, but
two of them can be determined as functions of the third one. This leads to 
some algebraic equations which ranks depend on the dimensionality 
of the brane. Somewhat special are the cases when the angular part blows up
and the spatial part goes to zero. If one assumes some relation between the
way in which the two of them vary towards these limits, this implies for 
$p=7,8,9$ some configurations in which the potential can have an
extremum, while for lower branes the
potential continues to have an infinite value. Nevertheless, when talking
about stable configuration, we see that the situation is different. For 
example if one consider a system that contains one antibrane and three 
parallel branes, the antibrane might be in equilibrium due to the fact that 
the forces on it from various branes may compensate each other, but its force 
on any brane cannot be compensated. Consequently, there is a collapse due
to the presence of three tachyonic fields in the system which in general 
interact among each other. If the tachyons are considered as non-interacting
fields, their potential is given by Eq.(\ref{tachpot}). However, in general 
case, its form is unknown. Due to the fact that the tachyons are off-shell
degrees of freedom, this potential should be obtained from an open string
field theory \cite{bsz,sz,as1}.

{\bf Acknowledgements} I acknowledge for useful discussions to
N.Berkovits and J. Schwarz. I would also like to thank to
C. A. T. Echevarria and C. R. dos Santos for suggestions and to
J. Abdalla Helay\"{e}l-Neto and M. A. de Andrade for supporting this
work at GFT-UCP. Also, this work was partially supported by a FAPESP
postdoc fellowship.

\vspace{2cm}


\end{document}